\documentclass[12pt]{iopart}
\usepackage[english]{babel}
\usepackage{graphicx,caption}
\usepackage[usenames,dvipsnames,svgnames,table]{xcolor}
\usepackage{CJK}
\newcommand{\bea}{\begin{eqnarray}}
\newcommand{\eea}{\end{eqnarray}}
\newcommand{\rvec}{\vec{r}}
\newcommand{\vvec}{\vec{v}}
\newcommand{\Fvec}{\vec{F}}
\newcommand{\xivec}{\vec{\xi}}
\newcommand{\f}{\displaystyle\frac}

\newcommand{\tick}{\textsc{\char13}}
\newcommand{\highlightRevision}{}
\begin{document}

\title[Spatial flocking: Control by speed, distance, noise and delay]{Spatial flocking:\\
  Control by speed, distance, noise and delay}

\author{Ill\'{e}s J. Farkas}
\address{Huazhong University of Science and Technology, 1037 Luoyu Rd, Wuhan, China, 430074}
\address{MTA-ELTE Statistical and Biological Physics Research Group, Hungarian Academy of Sciences, P\'{a}zm\'{a}ny P\'{e}ter s\'{e}t\'{a}ny 1A, Budapest, Hungary, 1117}
\ead{fij@elte.hu}

\begin{CJK}{UTF8}{gbsn}
\author{Shuohong Wang (王硕鸿)}
\end{CJK}
\address{School of Computer Science, Fudan University, 825 Zhangheng Rd, Shanghai, China, 201203}

%\begin{indented}
%\item[]\today
%\end{indented}

\begin{abstract}
  Fish, birds, insects and robots frequently swim or fly in
  groups. During their 3 dimensional collective motion, these agents
  do not stop, they avoid collisions by strong short-range repulsion,
  and achieve group cohesion by weak long-range attraction. In a
  minimal model that is isotropic, and continuous in both space and
  time, we demonstrate that (i) adjusting speed to a preferred value,
  combined with (ii) radial repulsion and an (iii) effective
  long-range attraction are sufficient for the stable ordering of
  autonomously moving agents in space. Our results imply that beyond
  these three rules ordering in space requires no further rules, for
  example, explicit velocity alignment, anisotropy of the interactions
  or the frequent reversal of the direction of motion, friction,
  elastic interactions, sticky surfaces, a viscous medium, or vertical
  separation that prefers interactions within horizontal layers. Noise
  and delays are inherent to the communication and decisions of all
  moving agents. Thus, next we investigate their effects on ordering
  in the model. First, we find that the amount of noise necessary for
  preventing the ordering of agents is not sufficient for destroying
  order. In other words, for realistic noise amplitudes the transition
  between order and disorder is rapid. Second, we demonstrate that
  ordering is more sensitive to displacements caused by delayed
  interactions than to uncorrelated noise (random errors). Third, we
  find that with changing interaction delays the ordered state
  disappears at roughly the same rate, whereas it emerges with
  different rates. In summary, we find that the model discussed here
  is simple enough to allow a fair understanding of the modeled
  phenomena, yet sufficiently detailed for the description and
  management of large flocks with noisy and delayed interactions. Our
  code is available at \verb"http://github.com/fij/floc".

\end{abstract}

\submitto{\JPD}
\maketitle

\section{Introduction: Collective motion in 2 and 3 dimensions}

In all fields of life recent technological developments have lead to a surge in data acquisition.
However, usually the obtained data can be put to practical use only with improved analytic and predictive methods.
For collective motion (swarming, active matter), some of the recent major experimental advances have been
the systematic measurements of fish trajectories in small shoals \cite{katz},
tracking the individual coordinates of up to 2700 birds in flocks \cite{ballerini_PNAS},
and obtaining GPS track logs of homing pigeons flying together \cite{nagy}.
Initially, experimental and modeling efforts were focused on planar (2 dimensional) motion.
Due to these efforts it is now well known
that in 2 dimensions bacteria, insects, horses, and also humans display collective motion patterns
\cite{czirok,buhl,fischhoff,helbing}.
Compared to planar motion, an agent moving in space can be kept aligned
by a higher number of nearest neighbor interactors.
At the same time, it has also more directions to turn away from the consensus of those nearest neighbors.

The most straightforward local rule that can describe
the alignment of moving agents with their neighbors 
is to set each agent's direction of motion explicitly 
to the average direction of its neighbors \cite{vicsek,jadbabaie}.
Turning continuously toward the average direction of the neighbors is also possible \cite{reynolds,tonertu}.
More detailed mechanisms of the alignment include
anisotropic interactions caused by elongated shapes \cite{narayan,zhang,ginelli,kaiser},
also combined with a frequent reversal of the direction of motion \cite{wu},
the preference for movements in the horizontal plane (as opposed to vertical movements) \cite{ballerini_AnimBeh},
a viscous medium \cite{bae},
friction among the agents and inelastic collisions \cite{schweitzer,grossman},
and sticking together \cite{mendelson}.
While these rules can set the direction of motion for the agents,
collision avoidance and cohesion (staying together) are also necessary for flock formation.
To avoid the collisions of moving and interacting agents
the simplest solution is to let all agents (in the model) have zero size \cite{vicsek}.
A more realistic solution is a strong short-range repulsive interaction,
in which the magnitude of the repulsion force becomes very high when two agents come too close.
Finally, for keeping the group together two commonly applied modeling tools are the spatial confinement of the group
(e.g., periodic boundaries)
and a weak attraction that is turned on when distances between the agents grow.

The model that we discuss here focuses on {\it controlling the speed of the agents individually}.
The speed of an agent is adjusted to the preferred speed
with a rate that is proportional to the difference from the preferred value.
This modeling approach is realistic, because
-- according to recent experimental and modeling evidence --
individual {\it speed control} plays a key role in the formation
and the stability of bird flocks and fish shoals \cite{katz,bialek,hemelrijk}.
Also, experiments and models for vibrated self-propelled hard disks have shown that
the binary collisions caused by {\it maintaining speed}
can align velocity vectors first locally,
and then also across the entire system \cite{deseigne}.

We investigate the effect of time delay and noise, too.
Time delay is a common phenomenon caused by latent communication
between agents, information processing cost, and inertial reasons \cite{olfati-saber,mier}.
Noise at all levels is also
inherent to the communication and decisions of all moving agents in a
dynamic system and can lead to transitions between behavioral patterns
\cite{mier,nguyen,ton,erban}.
Regarding the combination of time delay and noise,
simulation results in \cite{lindley} showed that a system with noise and
delay displays bistability of several coherent patterns.
Here we investigate both aspects and show their fundamental dissimilarities.

\section{Model: A minimal continuous description of spatial flocking}
\label{sec:model}

\begin{figure}
\centering
\captionsetup{width=\textwidth}
\includegraphics[width=\textwidth]{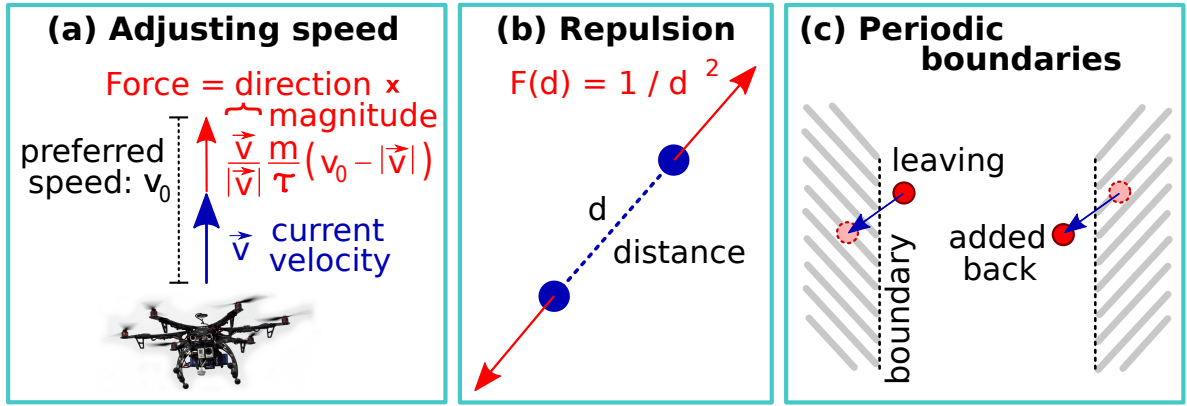}
\caption{The spatial flocking model used in the paper. Image from thedroneinfo.com.}
\label{fig:1}
\end{figure}

Here we investigate the 3-dimensional version of the generic model that was suggested in \cite{vicsek_and_zafeiris}
and analyzed for 2 dimensions in \cite{pre15}.
The model is continuous in both space and time, and it contains $N$ agents.
The $i$th agent's position is $\rvec_i(t)=(x_i(t),y_i(t),z_i(t))$
and its velocity is $\vvec_i(t)=(v_{x,i}(t),v_{y,i}(t),v_{z,i}(t))$.
During its motion each agent continuously adjusts its speed,
$v_i(t)$, toward a constant preferred $v_0$ value with characteristic time $\tau$.
To model collision avoidance, we apply pairwise radial repulsion among the two agents
with a magnitude of $F(r) = c\, r^{-2}$ as a function of their distance,
$r$ (for correct dimensions we set $c = 1 N m^2$).
For simplicity, we set the mass of each agent to $m=1kg$ and their time constants for adjusting speed to $\tau=1s$.
Based on the above, the equations of motion are (for $i=1\dots N$):
\bea
m \, \f{d \vvec_i}{dt} = \
\f{m}{\tau} \, \f{\vvec_i}{|\vvec_i|} \, \Big(v_0-|\vvec_i|\Big) - \sum_{j\not= i}\Fvec(|\rvec_i-\rvec_j|) \, .
\label{eq:1}
\eea
\noindent
Note that the first term on the right hand side of Eq.(\ref{eq:1})
points toward the $i$th agent's own direction of motion, $\vvec_i/|\vvec_i|$.
In other words, this simplified model separates collision avoidance (second term) from
keeping the preferred speed (first term).

We solve the equations of motion numerically by applying the midpoint method \highlightRevision{\cite{midpoint}}
with an integration time step of $\Delta t=10^{-3}s$
in a cube that has side length $L=50m$ and periodic boundary conditions in all three directions.
When the simulation is started we place all agents at random positions
-- but no pair of them is allowed to be closer than $0.6 \, L \, N^{-1/3}$ --
and set all speeds to the preferred speed, $v_0=5m/s$.
We start the system either from a disordered state or an ordered state.
When starting the system from the disordered state, we set the directions of the velocity vectors
to $N$ {\it different}, randomly selected directions at simulation time $t=0$.
When starting from the ordered state, we set all directions to the {\it same},
randomly selected direction at time $t_{0}=-10^5$,
then simulate the system until $t=0$,
and after that start to log positions and velocities.

The two additional aspects of the model that we investigate are its
responses to noise and time delays.
We include noise into the model by adding a random $\vec{\xi}$ vector to the right hand side of Eq.(\ref{eq:1}).
This vector is uncorrelated both in time and among agents,
its direction is distributed uniformly in space,
and its magnitude is a constant, $\xi$.
If $\xi=1$, then the autocorrelation of $\vec{\xi}$ is 1,
therefore, the noise vector added to the r.h.s. of Eq.(\ref{eq:1})
during a simulation update of length $\Delta t$ is $\xi\sqrt{\Delta t}$.
Similary, for the first half-step of the midpoint update the amplitude of the
applied noise vector is $\xi\sqrt{\Delta t/2}$.
In addition to the above, we compute interactions with a distance-based upper cutoff:
in the simulations two agents interact only if their distance is below a fixed cutoff radius, $R=10m$.
Finally, for efficient computation, we apply a 3-dimensional grid with side length $R$
and search for an agent's interactors -- i.e., other agents closer than $R$ --
only within its own grid cell and the neighboring 26 grid cells.

\begin{figure}
  \centering
  \includegraphics[width=0.75\textwidth]{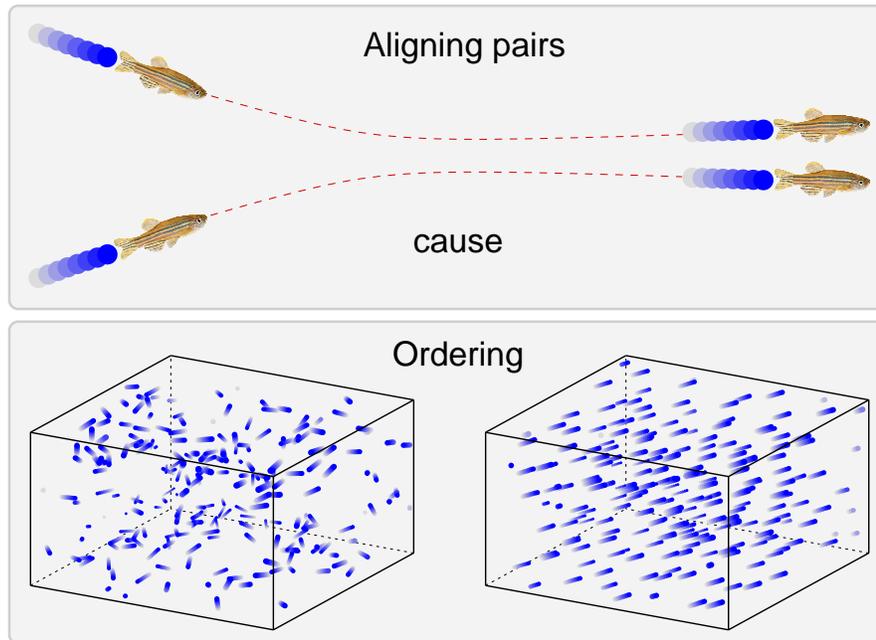}
  \caption{The alignment of pairwise interacting agents can cause large-scale ordering.
    {\bf Top panel.} In many cases the motion of two interacting agents
    (e.g., fish, birds or humans) becomes more parallel after their encounter.
    Here two fish are shown as an illustration, and their
    paths are indicated by sequences of colored filled disks.
    Image source: Wikipedia.
    {\bf Bottom panel.}
    Paths of $N=200$ agents for $\Delta t=0.4$ time in an actual simulation 
    -- midpoint integration of Eq.\,(\ref{eq:1}) --
    that was started from a disordered initial state (bottom left)
    and then reached the ordered state (bottom right).
  }
\label{fig:2}
\end{figure}

\section{Results: Alignment and flocking}

A necessary condition for the emergence of a single stable 
aligned group containing the majority of all agents is
the ability of small groups to align.
More specifically, the most frequently studied necessary condition
is whether the alignment of {\it two} agents increases during their encounter,
i.e., when they come close and depart (see top panel of Figure\,\ref{fig:2}).

\subsection{Alignment of two agents in a symmetric encounter}

\begin{figure}
  \centering
  \includegraphics[width=0.75\textwidth]{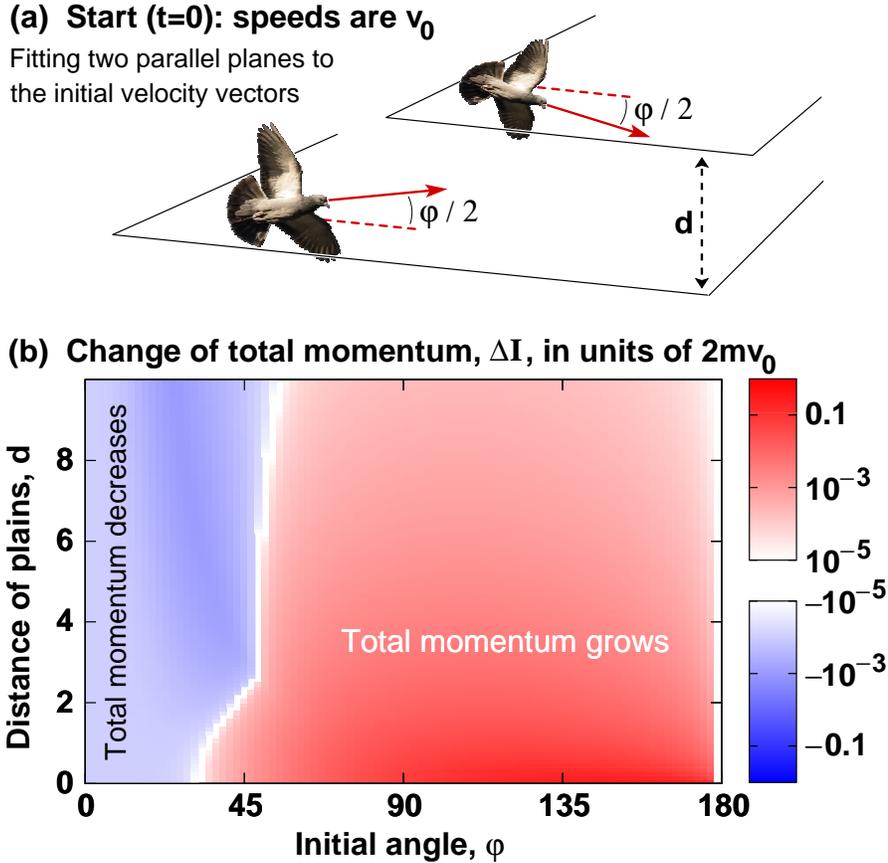}
  \caption{{\bf (a)} Initial state ($t=0$) of the two symmetrically moving identical agents.
  Image from openclipart.org.
  {\bf (b)} 
  Change of the total momentum of the two agents as a function of the initial angle, $\varphi$,
  and the distance, $d$, of the planes of their initial velocities.
  For the major part of the $(d,\varphi)$ parameter range
  the two agents' total momentum grows during their encounter,
  i.e., they become more aligned. See text for details.
}
\label{fig:3}
\end{figure}

As a simple -- yet nontrivial -- case of a two-agent encounter in space,
we analyse the scenario when two {\it identical}
agents move such that their velocity vectors are permanently mirror images of each other
with respect to the two agents' center of mass (Fig.\,\ref{fig:3}a).
To parametrize this 3-dimensional motion of the two agents,
we fit two parallel planes to the skew lines defined by their velocity vectors at $t=0$.
The distance of these two plains is $d$, and both agents have a speed of $v_0$.
When looking at the two agents from the direction perpendicular to the two planes,
at $t=0$ the distance of the two agents is $1,000m$ and the angle of their velocity vectors is $\varphi$.
As in Eq.\,(\ref{eq:1}), radial repulsion is $F(r) = c\, r^{-2}$ (with $c = 1 N m^2$ for correct dimensions).
We compute this pair encounter by integrating Eq.\,(\ref{eq:1}) with the
forward Euler method with an integration time step of $dt=10^{-5}$,
and the parameters $v_0=1m/s, \tau=1s$ and $\xi=0$.
We stop the integration when the horizontal distance of the agents reaches $1,100m$.
\highlightRevision{Finally, we obtain an estimate of the error
  of the numerical integration: we test how the integration time step parameter ($dt$)
  influences the calculated change
  of this two-particle system's total momentum ($\Delta I$) during the encounter.
We find that the maximal difference of the final $\Delta I$ values between
the $dt=10^{-4}$ and $dt=10^{-5}$ cases is $1.8\times 10^{-4}$.
This difference is significantly smaller than the $\Delta I$ values in
Figure\,\ref{fig:3} that are all above $10^{-3}$.}

According to Figure\,\ref{fig:3}b, we obtain the following result
for the investigated symmetric collision.
If the initial angle, $\varphi$, exceeds a threshold value
-- which is between $40$ and $50$ degrees depending on the distance,
$d$, of the two particles' initial planes, --
then the total momentum of the two particles increases.
In other words, Eq.\,(\ref{eq:1}) implies that 
two particles arriving symmetrically at a large angle
will depart at a smaller angle: they will become more parallel.
With many interacting and moving particles the
interactions are usually not pairwise and typically not symmetrical.
Therefore, the results displayed in Figure\,\ref{fig:3}b are not more than
a significant microscopic result pointing at the
possibility of the large-scale alignment of all moving agents.
This possibility is tested in Section\,\ref{subsec:ordering} below.

\subsection{Alignment of many collectively moving agents}
\label{subsec:ordering}

\begin{figure}
  \centering
  \includegraphics[angle=-90,width=0.75\textwidth]{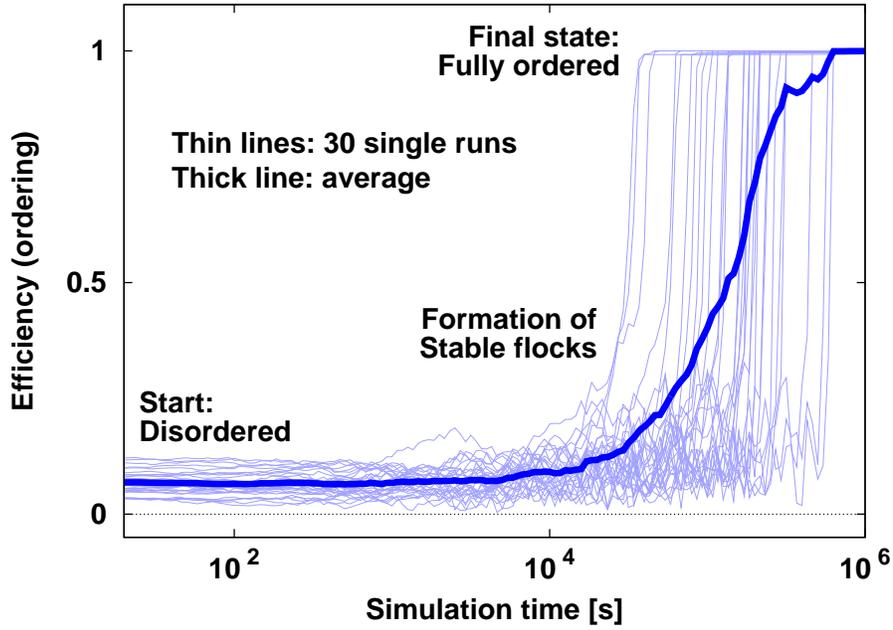}
  \caption{Transition from the initial disordered state (low $E$)
   to the ordered state (high $E$) in the model
   of Eq.\,(\ref{eq:1}) without noise ($\xi=0$) \cite{noise}.
   See text for details.
}
\label{fig:4}
\end{figure}

To test whether Eq.\,(\ref{eq:1}) can indeed lead to a stable coherent motion of all
agents, we start the system from a fully disordered setup
and check if it reaches a stable ordered state (Fig.\,\ref{fig:4}). 
We simulate the motion of agents
-- i.e., perform the midpoint integration of Eq.\,(\ref{eq:1}) --
within a cube that has side length $L=50$ and periodic
boundary conditions.
Parameters are
$v_0=5$ (preferred speed), $\tau=1$ (time constant for reaching the preferred speed), $R=10$ (interaction cutoff radius),
$dt=10^{-3}$ (time step of integration)
and $\xi=0$ (no noise).
At simulation time $t=0$ the speed of each agent is $v_0$ and velocity
vectors point to independently selected random directions.
Starting the system with this scenario, we measure
overall ordering through the quantity called {\it efficiency} \cite{eff-one}:  
\bea
E(t) = \f{1}{N v_0} \, \Bigg| \sum_{i=1}^{N} \vvec_i(t) \Bigg| \, .
\eea
\noindent
For $N=200$ agents Figure\,\ref{fig:4} shows the efficiency in $30$ single simulation runs
and also the average over all $30$ runs.
Observe in this figure that the formation of small stable flocks
(slightly elevated values of the efficiency compared to the initial level)
is quickly followed by the stable ordering of all agents ($E$ moves quickly up close to 1).
In other words, for an individual system the transition from disorder to order is {\it fast},
and the slowness of the ordering (growth) seen on the averaged curve is caused merely by the wide temporal
distribution of the fast transitions of the individual systems.

\begin{figure}
  \centering
  \includegraphics[angle=-90,width=0.8\textwidth]{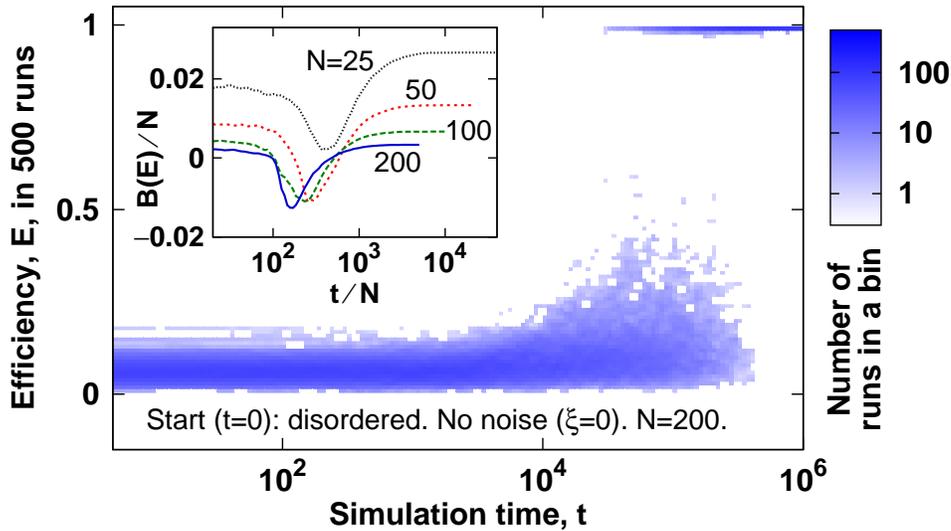}
  \caption{ {\bf Main panel.}
   Ordering of $N=200$ particles started from the
   disordered state $500$ times randomly.
   Pixel color depths show the number of independent simulation runs
   that have an efficiency within a given interval at the given simulation time point.
   The ratio of adjacent time values is $10^{1/30}$,
   and the width of $E$ intervals is $0.01$.
   {\bf Inset.} The normalized Binder cumulant, $B(E)/N$,
   of the distribution of the $500$ values of $E(t)$.
   shows that the transition remains bimodal (i.e., discontinuous) with increasing system size.
   With changing $N$ we change $L$ also to keep
   the density of particles -- $N \times L^{-1/3}$ -- constant.
   See text for the definition of $B$ and further details.
}
\label{fig:5}
\end{figure}

Figure\,\ref{fig:5} shows an alternative view of the same transition with a much larger data set.
Here we evaluate $500$ simulation runs (instead of the previous $30$) to obtain
the distribution of these $500$ efficiency values at each of the selected time points.
Similar to Figure\,\ref{fig:4},   
we observe on the main panel of Figure\,\ref{fig:5} 
that between the simulation times $t=10^4$ and $10^6$
each system moves relatively quickly from the
\highlightRevision{disordered state ($0\le E\ll 1$)} to the ordered state ($E\approx 1$).
Regarding transitions, a frequently investigated question is
how the type of the transition changes with increasing system size \cite{nagy07,aldana}.
In the inset of Figure\,\ref{fig:5} we quantify the distribution of efficiency values at the selected 
simulation time points by computing the Binder cumulant, $B(E)$, of the numerically measured $E$ values:
\bea
B(E)=\f{1-\langle E^4\rangle}{3 \langle E^2\rangle^2} \, .
\eea
We find that growing system size (and constant density)
the transition from the ordered to the disordered state remains rapid.
For practical applications this means that 
-- in the absence of disturbances --
once the transition to the ordered state 
starts, it is very hard to stop, and the resulting ordered state is very stable.
To go into more detail,
in the following we investigate the stability of the arising ordered state
to two of the most common disturbances: noise (errors) and delay (also called time lag).

\subsection{Stability of the ordered state to noise}
\label{subsec:noise}

\begin{figure}
  \centering
  \includegraphics[angle=-90,width=0.75\textwidth]{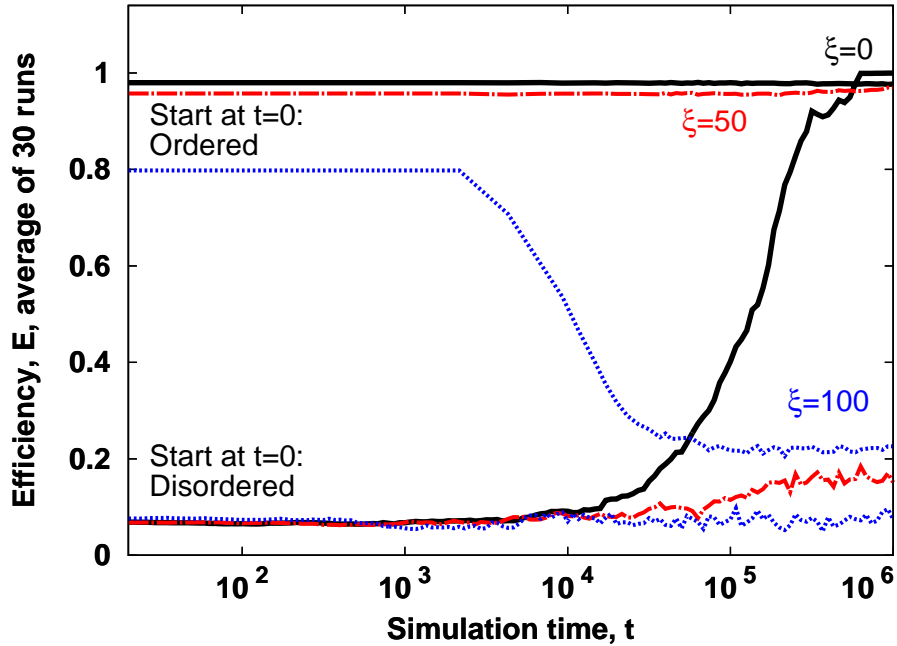}
  \caption{According to the current model,
    the lowest amount of noise that can keep the system disordered 
    is not sufficient for destroying order.
   This property -- called hysteresis -- makes order very stable
   even at intermediate noise levels.
   On the other hand, it also implies that a transition started at intermediate noise levels is very fast.
   Parameters are as in Section\,\ref{subsec:ordering}.
   Here the ordered state was measured after running (equilibrating) the simulation for $10^5$s.
   See text for details.
}
\label{fig:6}
\end{figure}

Errors and small changes (also called fluctuations or noise) are ubiquitous
in natural and social phenomena.
In most cases, a low amount of noise still allows ordering,
however, when noise amplitudes grow very large,
order is destroyed and the system becomes disordered \cite{freezing}.
In Figure\,\ref{fig:6} we test whether it is easy to destroy the ordered state (high $E$)
of the agents of Eq.\,(\ref{eq:1}) by increasing the initially low amount of noise to a high value.
With the parameters selected in Section\,\ref{subsec:ordering} we find that
the noise level $\xi=50$ can already keep an initially disordered system from reaching order,
however, it is not sufficient to destroy order in an initially ordered system.
We conclude that the system shows hysteresis,
which is similar to the 2-dimensional version of the model \cite{pre15}.
Finally, note that this behavior stabilizes the ordered state,
and at the same time makes transitions fast.

In this final paragraph of Section\,\ref{subsec:noise},
we add a note regarding how we start the system from the ordered state.
For the curves labeled ``Start: ordered'' in Figure\,\ref{fig:6}
we start each of the 30 simulations (with 30 different random seeds)
at $t_{\mathrm{start}}=-100,000s$ by setting the velocity vectors of the agents
fully aligned and their coordinates randomly with a minimal distance.
Here minimal initial distance means
-- similar to the case of the disordered initialization in Sec.\,\ref{sec:model} --
a lower bound of $0.6 \, L \, N^{-1/3}$ for the agent-agent distance.
First, starting at $t_{\mathrm{start}}$, we run the simulation until $t=0s$ to reach spatial ordering.
After the equilibration interval that starts at $t_{\mathrm{start}}$ and ends at $t_0$,
we run the numerical integration of the equations of motion until $t=10^5$ or $t=10^6$.
Note that in Sec.\,\ref{subsec:delay} we do not apply the equilibration technique
described here.

\subsection{Stability of the ordered state to delay (time lag)}
\label{subsec:delay}

\begin{figure}
  \centering
  \includegraphics[angle=-90,width=0.75\textwidth]{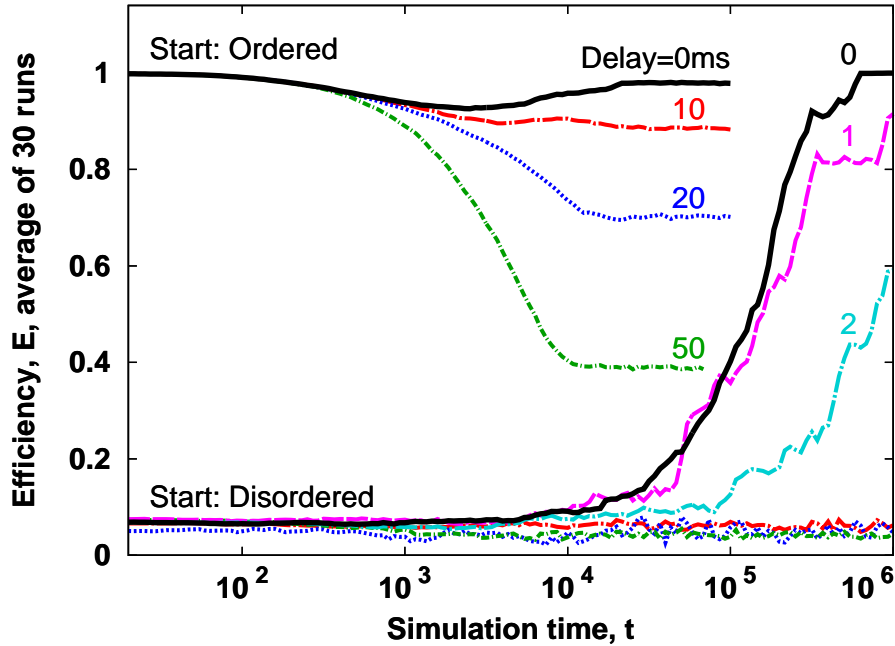}
  \caption{Time delays (time lags) introduced to the interactions of the model
           affect the ordered and the disordered states differently.
           During the emergence of order, short time lags (even $1$ or $2$ ms)
           can significantly change the rate of ordering.
           On the other hand, for highly diverse values (up to $50$ms) of the time lag 
           order is destroyed on the same time scale.
           The time delay values of the interactions are shown in milliseconds
           over the respective measured average curves.
           Note that at the start of the simulation the efficiency drops slightly and then grows again to reach $E\approx 1$.
           To explain this effect, recall that this simulation is started with directionally ordered agents
           that are spatially not yet ordered.
           The slight drop of the efficiency shows as the particles attain their spatial order (close to crystalline ordering of coordinates)
           while temporarily losing some of their directional alignment.
}
\label{fig:7}
\end{figure}

In most natural, social and technological phenomena the delays of interactions 
play a crucial role in shaping the emerging collective behavioural patterns.
Stable formations of collective motion can usually emerge only if 
the distance traveled by an agent 
over the delay time interval 
with its characteristic relative speed (relative to nearby agents and obstacles)
remains safely below the agent's distance to those other agents and obstacles.
For the current model we set a single delay time interval, $t_d$, that affects
in an identical way how an agent responds to any internal or external effect.
For the numerical integration, we implement delayed interactions by setting
the time lag, $t_d$, of all effects (forces) 
to a multiple of the numerical integration time step, $\Delta t$.
Next, we replace
all forces on the right hand side of Eq.\,(\ref{eq:1}) by the 
values of the same forces measured $t_d$ time before the current time.
Note that these forces include self-propelling, therefore,
a particle's response to changes in its own speed
are delayed as well.

We find that -- similar to noise --
interaction delays act differently on the ordered and disordered states of the system (see Fig\,\ref{fig:7}).
First, the two states respond to time delays at different time scales.
Whereas the ordered state responds to a wide range of
interaction delays on the same short time scale,
the already slow disorder $\rightarrow$ order transition is further significantly
delayed even by small amounts of the interaction delay.
Second,
even large amounts (up to $t_d=50$ms) of the interaction delay cannot 
entirely destroy the ordered state.
Third, based on our current results with $t_d=0$, $10$, $20$ and $50$,
we conclude that ordering is lost as a linear function of the amount of interaction time lag, $t_d$.

\section{Discussion and Outlook}

In this paper, we investigated a minimal continuous model of
3-dimensional collective motion.
The model contains continuous adjustment of particle speed to a
preferred value, pairwise radial repulsion for collision avoidance,
and an effective weak attraction (periodic boundaries).
We found that the combination of these three model components is sufficient
for stable spatial ordering, and beyond these three no further
model components are necessary.
After investigating the model on the microscopic level,
we found that for the majority of symmetric two-agent
encounters the total momentum of the two agents increases.
Regarding the macroscopic level,
we found that the transition from disorder to order is fast for
both small and large system sizes,
which is in good agreement with previous results \cite{nguyen}.
In the minimal continuous model that we investigated we found also that 
if the noise intensity is above a threshold value, then the
system cannot reach the ordered state, similarly to results reported in
\cite{romenskyy}.

\section{Acknowledgements}

The authors gratefully acknowledge support from the Hungarian Scientific Research Fund (OTKA NN 103114),
and advice received from T. Vicsek, M. Nagy, and H.-T. Zhang.


\begin{thebibliography}{9}
\bibitem{midpoint}
  \highlightRevision{For two particles the precision of the forward Euler method is sufficient (Fig.~\ref{fig:3}).
    To integrate the equations of motion of many particles,
    we apply the midpoint method (Fig.~\ref{fig:4} and later figures).}
\bibitem{eff-one}
Note that Eq.\,(\ref{eq:1}) does allow $E(t)$ to exceed $1$ 
-- if many agents have $|\vvec_i|>1$, --
however, even with $2$ agents this is a very rare case.
\bibitem{noise}
Even without explicit noise
(i.e., if in Eq.\,(\ref{eq:1}) the noise amplitudes are $\xi(t)=|\xivec_i(t)|=0$)
there are two sources of noise in the simulations:
the errors of the numerical integration method and rounding.
Usually (as in the present case) the numerical integration method
is the dominant source of noise.
\end{thebibliography}
\end{document}